\documentstyle[twoside,fleqn,espcrc2]{article}
\def\etall{{\it et. al. }}
\def\etal,{{\it et. al., }}
\def\huntsf{ {\it The 4th Huntsville Meeting}.}

\def\g{\gamma}

\def\E051{(E/10^{51}{\rm ergs})}
\def\R07{(R_i/10^7{\rm cm})}

\begin{document}
\title{Gamma-Ray Bursts and Related Phenomena}
\author{Tsvi Piran \\
Racah Institute for Physics, The Hebrew University, \\ Jerusalem,
Israel 91904}

\begin{abstract}
  Gamma-ray bursts (GRBs) have puzzled astronomers since their
  accidental discovery in the sixties. The BATSE detector on the
  COMPTON-GRO satellite has been detecting one burst per day for the
  last six years. Its findings have revolutionized our ideas about the
  nature of these objects. They have shown that GRBs are at
  cosmological distances. This idea was accepted with difficulties at
  first. However, the recent discovery of an x-ray afterglow by the
  Italian/Dutch satellite BeppoSAX led to a detection of high
  red-shift absorption lines in the optical afterglow of GRB970508 and
  to a confirmation of its cosmological origin.  The simplest and
  practically inevitable interpretation of these observations is that
  GRBs result from the conversion of the kinetic energy of
  ultra-relativistic particles flux to radiation in an optically thin
  region. The ``inner engine" that accelerates the particles or
  generates the Poynting flux is hidden from direct observations.
  Recent studies suggest the {\bf ``internal-external'' model:} {\it
    internal shocks that take place within the relativistic flow
    produce the GRB while the subsequent interaction of the flow with
    the external medium produce the afterglow}. The ``inner engine''
  that produces the flow is, however, hidden from direct observations.
  We review this model with a specific emphasis on its implications to
  underground physics.
\end{abstract}
\maketitle

\section{Introduction}
\label{sec:intro}

Gamma-ray bursts (GRBs), short and intense bursts of $\sim
100$keV-1MeV photons, were discovered accidentally in the sixties by
the Vela satellites\cite{Kle}. The mission of the satellites was to
monitor the ``outer space treaty" that forbade nuclear explosions in
space.  A wonderful by-product of this effort was the discovery of
GRBs.  Had the satellites not been needed for security purposes, it is
most likely that today we would still be unaware of the existence of
these mysterious bursts.

The discovery of GRBs was announced in 1973 \cite{Kle}.  Since then,
several dedicated satellites have been launched to observe the bursts
and numerous theories were put forward to explain their origin.  In
the mid eighties a consensus formed that GRBs originate from Galactic
neutron stars.  The BATSE detector on the COMPTON-GRO (Gamma-Ray
Observatory) was launched in the spring of 1991.  It has
revolutionized GRB observations and consequently our basic ideas on
their nature. It ruled out the galactic disk neutron star model.
While BATSE's observations could not rule out the possibility that
GRBs originate from objects in the extended galactic halo the
observations strongly suggested that the sources of GRBs are
extra-galactic at cosmological distances. This idea was recently
confirmed by the discovery by BeppoSAX \cite{Costa97a} of an X-ray
transient counterparts to several GRBs which was followed by the
discovery of optical \cite{vanParadijs97a,Bond97} and radio transients
\cite{Frail97a}.  Absorption lines with $z = 0.835$ were measured in
the optical spectrum of the counterpart to GRB970508,
\cite{Metzger97a}, providing a lower limit to the redshift of the
optical transient and the associated GRB.

The cosmological origin of GRBs implies that GRB sources are much more
luminous than previously thought. They release $\sim 10^{51}-
10^{52}$ergs in a few seconds putting them as the most
(electromagnetically) luminous objects in the Universe. This also
implies that GRBs are rare events.  BATSE observes on average one
burst per day and this corresponds to a rate of about one burst per
million years per galaxy \cite{Piran92}.

A generic scheme of a cosmological GRB model has emerged in the last
few years \cite{Piran97}.  The recently observed x-ray, optical and
radio counterparts were predicted by this picture
\cite{PacRho93,Katz94,SaP97a,MR97}. This discovery provides 
a confirmation of this model
\cite{Wiejers_MR97,Waxman97a,Vietri97,KP97,MesReesWei97}.  
According to this scheme the observed $\gamma$-rays are emitted when
an ultra-relativistic energy flow is converted to radiation. Possible
forms of the energy flow are kinetic energy of ultra-relativistic
particles or electromagnetic Poynting flux. This energy is converted
to radiation in an optically thin region, as the observed bursts are
not thermal.  The energy conversion occurs either due to the
interaction with an external medium, like the ISM \cite{MR1} or due to
internal process, such as internal shocks and collisions within the
flow \cite{NPP,MR4}.  Recent work \cite{SaP97a,PiSa97} shows that the
external shock scenario is quite unlikely, unless the energy flow is
confined to an extremely narrow beam or the process is highly
inefficient. The alternative is that the burst is produced by internal
shocks.  Not all the energy is converted to radiation in these shocks
\cite{MMM95,KPS97}.  The remaining energy is converted to radiation in
subsequent external shocks producing the afterglow \cite{SaP97a}. We
call this the {\bf internal-external} shock model.

The ``inner engine" that produces the relativistic energy flow is
hidden from direct observations. However, the observed temporal
structure seen in the bursts reflects directly this ``inner engine's''
activity \cite{KPS97}.  This model requires a compact inner engine
that produces an irregular  ``wind'' -- a long energy flow (long
compared to the
size of the engine itself) -- rather than an explosive engine that
produces a fireball whose size is comparable to the size of the
engine.

At present there is no agreement on the nature of the ``engine'' -
even though binary neutron star mergers \cite{Eichler89} are a
promising candidate. All that can be said with some certainty is that
whatever drives a GRB must satisfy the following general features: (i)
It should produce an extremely relativistic energy flow containing
$\approx 10^{51}-10^{52}$ergs. (ii) The flow is highly variable and it
should last for the duration of the burst (typically a few dozen
seconds). It may continue at a lower level on a time scale of a day or
so \cite{KaPS97}.  (iii) Finally, it should be a rare event occurring
about once per million years in a galaxy. The rate is, of course,
higher and the corresponding energy is lower if there is a significant
beaming of the gamma-ray emission.

\section{Observations}

\label{sec:obs}

GRBs are short, non-thermal bursts of low energy $\gamma$-rays.  It is
quite difficult to summarize their basic features.  This difficulty
stems from the enormous variety displayed by the bursts.  A
``typical'' GRB (if there is such a thing) lasts for about 10sec.
However, the observed durations vary by five orders of magnitude, from
several milliseconds \cite{Fishman93}, to $10^3$sec
\cite{Klebedasel84}.  In one case high energy (GeV) photons were
observed several hours after the main pulse \cite{Hurley94}.  The
bursts have  complicated and irregular time profiles which varies
drastically from one burst to another.  In most bursts, the typical
variation takes place on a time-scale, $\delta T$ significantly
smaller than the total duration of the burst, $T$. We denote the ratio
by ${\cal N} = T/\delta T$ and typically ${\cal N} \approx 100$.

GRBs are characterized by emission at the few hundred keV ranges with
a non-thermal spectrum.  Most bursts are accompanied by a high energy
tail which contains a significant amount of energy -- $E^2 N(E)$ is
almost a constant.  Several bursts display high energy tails up to 26
GeV\cite{Hurley94}.  In fact EGRET and COMPTEL (which are sensitive to
higher energy emission but have higher thresholds and smaller fields
of view) observations are consistent with the possibility that all
bursts have high energy tails \cite{Dingus97}.  The high energy tails
lead to a strong constraint on GRB models. The high energy photons
must escape freely from the source without producing electron positron
pairs!  As we see in section \ref{sec:compact} this provides the first and
most important clue on the nature of GRBs.  

GRB observations were revolutionized on February 28 1997 with
discovery of an X-ray counterpart to GRB970228 by the Italian-Dutch
satellite BeppoSAX \cite{Costa97a}.  The accurate position determined
by BeppoSAX enabled the identification of an optical afterglow
\cite{vanParadijs97a} - a decaying point  source surrounded by a 
red nebulae.  Following observations with HST \cite{Sahu97} revealed
that the nebula is roughly circular with a diameter of 0''.8. The
nebula's intensity does not vary, while the point source decays with a
power law index $\approx {-1.2}$ \cite{Galma97a}.  X-ray observations
by BeppoSAX, ROSAT and ASCA revealed a decaying x-ray flux $\propto
t^{-1.33\pm0.11}$.  The decaying flux can be extrapolated as a power
law directly to the x-ray flux of the second peak (even though this
extrapolation requires some care in determining when is $t=0$).

Afterglow was also detected from GRB970508.  This $\gamma$-ray burst
lasted for $\sim 15$sec, with a $\gamma$-ray fluence of $\sim 3
\times 10^{-6}$ergs/cm$^{-2}$. Variable emission in x-rays, optical
\cite{Bond97} and radio \cite{Frail97a} followed the
$\gamma$-rays. The spectrum of the optical transient revealed a set of
absorption lines associated with Fe II and Mg II with a redshift
$z=0.835$ \cite{Metzger97a}. A second absorption line system with
$z=0.767$ is also seen. In addition there are O II emission lines with
a redshift $z=0.835$. This sets the cosmological redshift of GRB970508
to be greater or equal than $0.835$.  The lack of Lyman alpha
absorption lines sets an upper limit of $z=2.1$ to this redshift.
The optical light curves show a clear peak at around 2 days after the
burst. After that it shows a continuous power law decay $\propto
t^{-1.18}$ \cite{Sokolov97}. Radio emission was observed first one
week after the burst \cite{Frail97a}. This emission showed intensive
oscillations which were interpreted as scintilations
\cite{Goodman97}. The subsequent disappearance of these oscillations 
after about three weeks enables Frail and Kulkarni \cite{Frail97a}
to estimate the size of the fireball at this stage to be $\sim 10^{17}$cm.
This was supported by the indication that the radio emission was
initially optically thick \cite{Frail97a}, which yields a similar
estimate to the size \cite{KP97}.

\section{Compactness,  Relativistic Motion and the 
Fireball Model.}
\label{sec:compact}

The key to understanding GRBs lies, I believe, in understanding how
GRBs bypass the compactness problem.  Consider a typical burst with a
total energy of $10^{51}$ergs (as inferred from the observed flux and
the implied distance of a cosmological source) that varies on a time
scale $\delta T \approx 10$msec.  Standard considerations suggest that
the temporal variability implies that the sources are compact with a
size, $R_i < c \delta T \approx 3000$km.  The observed spectrum
contains a large fraction of the high energy $\gamma$-ray photons.
These photons could interact with lower energy photons and produce
electron-positron pairs via $\gamma \gamma \rightarrow e^+ e^- $.  The
average optical depth for this process is $\sim 10^{15}(E/10^{51}{\rm
ergs}) (\delta T / 10~{\rm msec})^{-2}$ \cite{Piran97}. However, the
observed non-thermal spectrum indicates with certainty that the source
must be optically thin.

The compactness problem can be resolved if the emitting region is
moving towards us with a relativistic velocity characterized by a
Lorentz factor, $\gamma \gg 1 $.  We detect blue-shifted photons whose
energy at the source is lower by a factor $\gamma$.  Fewer photons
have sufficient energy to produce pairs.  Additionally, relativistic
effects allow the radius from which the radiation is emitted to be
larger than the previous estimate by a factor of $\g ^2$: $R_e \le
\gamma^2 c \delta T$.  The resulting optical depth is lower by a
factor $ \g ^{(4+2\alpha)}$ (where $\alpha \sim 2$ is the spectral
index). The compactness problem can be resolved if the sources are
moving relativistically towards us with Lorentz factors $\g >
10^{15/(4+2\alpha)} \approx 10^2$. It must be stressed that the motion
is not necessarily pointed towards us. While we might be looking at a
jet pointing towards us it is also possible that the motion is
spherically symmetrically outwards away from some center.

The potential of relativistic motion to resolve the compactness
problem was realized in the eighties by Goodman \cite{Goo86},
Paczy\'nski \cite{Pac86} and Krolik \& Pier \cite{KroPie}.  While
Krolik \& Pier \cite{KroPie} considered a kinematical solution, Goodman
\cite{Goo86} and Paczy\'nski \cite{Pac86} considered a dynamical
solution in which the relativistic motion results naturally when a
large amount of energy is released within a small volume. They show
that this would result in a relativistic explosion, which is called a
fireball.  Goodman \cite{Goo86} and Paczy\'nski \cite{Pac86}
considered  pure radiation fireballs. Shemi \& Piran \cite{SP}
have shown  that if the fireball contains baryonic mass it will
become relativistic only if the initial rest mass energy, $Mc^2$, is
small compared to the total energy $E$. In these cases the initial
energy of this fireball will be converted to the kinetic energy of the
baryons, whose Lorentz factor is simply $\gamma = E/M c^2$.

The kinetic energy is converted to ``thermal'' energy of relativistic
particles via shocks.  Both the low energy spectrum of GRBs and the
high energy spectrum of the afterglow provide indirect evidence for
relativistic shocks in the GRB \cite{Cohen97} and in the afterglow
\cite{Wiejers_MR97}.  There are two modes of energy conversion (i)
External shocks, which are due to interaction with an external medium
like the ISM \cite{MR1}. (ii) Internal shocks that arise due to shocks
within the flow when fast moving particles catch up with slower ones
\cite{NPP,MR4}. In either case these shocks must take place at 
sufficiently large radii where the flow is optically thin, allowing
the emission of a non-thermal spectrum.

\section{The Angular Spreading Problem}

External shocks are practically inevitable if the fireball is
surrounded by some external medium, such as the ISM.  Internal shocks
are more demanding. They require that the flow will be irregular and
will contain faster shells that will catch up with slower ones. 
External shocks were considered, therefore,  as the canonical model
while internal shocks were considered as a more exotic
variant. However, Sari \& Piran \cite{SaP97a} have recently shown that
external shocks cannot produce the complicated highly variable
temporal structure observed in most GRBs.

Let $\Delta $ be the width of the shell and let the energy conversion take
place between $R_{E}$ and $2R_{E}$. The emitting material moves with a
Lorenz factor, $\gamma_e $. There are three generic time scales.  (i)
The radial time scale, $T_{R}$: The difference in arrival time between
two photons emitted at $R_{E}$ and $2R_{E} $ - $T_{R}\approx
R_{E}/\gamma_e ^{2}c.$ (ii) The angular time scale, $T_{angular}$: The
difference in arrival time between two photons emitted along the line
of sight and at an angle $\theta$ from the line of sight. Because of
relativistic beaming an observer detects radiation from an angular
scale $\gamma_e^{-1}$ around the line of sight.  Thus, the angular
size of the observed regions always satisfies $\theta \le
\gamma_e^{-1}$ and $T_{angular }\approx R_{E}\theta^{2}/c \le
R_{E}/\gamma_e^{2}c$.  (iii) The shell crossing time, $T_\Delta$: The
light crossing time of the shell corresponds to the time difference
between the photons emitted from the shell's front and from its
back. This equals: $ T_\Delta =\Delta /c.$ Quite generally a forth
time scale, the cooling time scale, is shorter than all those scales.

Comparison of $T_{R}$ and $T_{angular}$ reveals that if the system is
``spherical'' ($\theta > \gamma_e ^{-1}$) then due to relativistic
beaming we have effectively $\theta \approx \gamma_e ^{-1}$ and $T_{R}
\approx T_{angular}$ \cite{Piran97,SaP97a,Fen}.   This leads to the 
angular spreading problem.  Blending of emission from regions from an
angle $\gamma_e^{-1}$ from the line of sight leads to smoothing of the
signal on a time scale: $T_{angulr} \approx T_R$.  Therefore, unless
$T_{\Delta}> T_{R}\approx T_{angular}$ there will be a smooth single
peak burst with $\delta T\approx T$.  It turns out that this will
always be the case if the emission is due to external shocks
\cite{SaP97a}.

One must break the spherical symmetry on scales smaller than $\gamma_e
^{-1}$ to produce a variable burst with $\delta T\ll T$ via an
external shock.  The angular size of the emitting regions must be
smaller than $(\gamma_e {\cal N})^{-1}\le 10^{-4}$ \cite{SaP97a}.  A
sufficiently narrow jet can satisfy this condition. However, it is not
clear how can such a narrow jet form. Furthermore, such narrow jets
are not observed elsewhere. Emission from numerous small size regions
must be highly inefficient in converting the kinetic energy to
radiation \cite{SaP97a} if we demand that the emitting regions are
sparse enough to produce the observed temporal variability.

Internal shocks would take place if the ``inner engine'' produces an
irregular wind (emission on a time scale much longer than the light
crossing time of the source).  These internal shocks could produce the
GRB.  Internal shocks take place at $R_E \approx \delta \gamma^2$
(where $\delta$ is the length scale of variability of the wind -
$\delta \le \Delta$ and $\gamma$ is the initial Lorentz factor). For
internal shocks the condition $T_{angular} \approx T_{R}<T_\Delta =
\Delta /c$ is always satisfied.  This will produce a burst whose
overall duration is $\Delta /c$ and the observed variability scale
is\footnote{ This is provided, of course, that the cooling time is
shorter than $T_{angular}$ \cite{SaP97b}.} $\delta T = \delta /c
\approx T_{angular }\approx T_{R}$. The variability scale could be
much shorter than the duration. The duration is determined  by the
activity of the inner engine and not by the emitting regions. The
observed temporal structure reflects the activity of the inner engine,
which must be producing a relatively long and highly irregular wind.
Numerical simulations of internal shocks can actually reproduce the
temporal structure observed in GRBs
\cite{KPS97}. 

\section{The Internal-External Model.}

Internal shocks can convert only a fraction of the total energy to
radiation \cite{MMM95,KPS97}.  A few month before the discovery of the
afterglow by BeppoSAX Sari \& Piran \cite{SaP97a} have pointed out
that after the flow has produced a GRB via internal shocks it will
interact via an external shock with the surrounding medium.  This
shock will produce the afterglow - a signal that will follow the GRB.
The idea of an afterglow in other wavelengths was suggested earlier
\cite{PacRho93,Katz94,MR97} but it was suggested as a follow up of
the, then standard, external shock scenario. In this case the
afterglow would have been a direct continuation of the GRB activity
and its properties would have scaled directly to the properties of the
GRB.  

According to internal-external model (internal shocks for the GRB and
external shocks for the afterglow)  different mechanisms produce the
GRB and the afterglow.  Therefore the afterglow should not be scaled
directly to the properties of the GRB.  This was in fact seen in the
recent afterglow observations.  In all models of external shocks the
observed time satisfy $t \propto R /\gamma_e^2$ and the typical
frequency satisfy $\nu \propto \gamma_e^4$. Since most of the emission
takes place at practically the same radius and all that we see is the
variation of the Lorentz factor we expect quite generally \cite{KP97}:
$\nu \propto t^{2 \pm \epsilon} $.  The small parameter $\epsilon$
reflects the variation of the radius and it depends on the specific
assumptions made in the model. We would expect that $t_x /t_\gamma
\sim 50 $ and $t_{opt}/t_\gamma \sim 300$.  The observations of
GRB970508 show that $(t_{opt}/t_\gamma)_{observed} \approx 10^4$. This
is in a clear disagreement with the single external shock model for
both the GRB and the afterglow.

\section{Afterglow Models}

Afterglow observations agree qualitatively with the synchrotron
cooling from a slowing down relativistic shell model
\cite{Wiejers_MR97,Waxman97a,Vietri97,KP97,MesReesWei97}.  
In all these models the shell is expanding, collecting more external
matter and slowing down. The Lorentz factor of the shell decreases and
this leads to a decrease in the typical synchrotron frequency.  The
shock front accelerated the electrons to some power law distribution
and these electrons cool via synchrotron (or Inverse Compton)
emission.  There are several variants to the basic model. These
include adiabatic vs.  radiative hydrodynamics, fast vs. slow cooling
of the shock heated electrons and synchrotron vs. synchrotron-self
Compton emission. Not all combinations are self consistent. For
example, radiative hydrodynamics occurs if the energy extracted by the
radiating electrons influences the hydrodynamics evolution of the
shell. Clearly, radiative hydrodynamics is incompatible with slow
cooling in which the electrons cooling time scale is longer than the
hydrodynamics time scale. So far there is no single clear model that
fits quantitatively all the afterglow data.

\section{The ``Inner Engine''}

The fireball model is based on an ``inner engine'' that supplies the
energy and accelerate the baryons.  This ``engine" is well hidden from
direct observations and it is impossible to determine what is it from
current observations. Unfortunately, the discovery of afterglow does
not shed an additional light on this issue.  For a long time the only
direct clues that existed on the nature of the ``inner engine'' were
the rate and the energy output.  It should be active at a rate of
about one per $10^6$years per galaxy, as this is the observed rate of
GRBs \cite{Piran92} and it should be capable of generating $\sim
10^{52}$ergs. Even these limits are not strict as an uncertainty in
the beaming angle, $\theta$, of the bursts leads to an uncertainty of
order $4 \pi /\theta^2$ in the rate and in the total energy involved.

The inner engine should be also capable of accelerating $\sim 10^{-7}
m_\odot$ to relativistic energies.  The need to convert the energy to
a relativistic flow is rather difficult to fulfill as it requires a
``clean'' system with a very low baryonic load.

The recent realization that energy conversion process is most likely
via internal shock rather than via external shocks provides additional
information about the inner engine.  According to this model the
relativistic flow must be irregular (to produce the internal shocks),
it must be variable on a short time scale (as this time scale is seen
in the variability of the bursts), and it must be active for up to a
few hundred seconds - as this duration determines the observed
duration of the burst. These requirements rule out all explosive
models.  The engine must be compact ($\sim 10^7$cm) to produce the
observed variability and it must operate for a few hundred seconds
(million times larger than the light crossing time) to produce a few
hundred seconds signals.

\section{Neutron Star Mergers}

Binary neutron star mergers (NS$^2$Ms) \cite{Eichler89} (or with a
small variant: neutron star-black hole mergers)  are, in my
mind the best candidate for the ``inner engine''.  These mergers take
place because of the decay of the binary orbits due to gravitational
radiation emission.  Pulsar observations suggest that NS$^2$Ms take
place at a rate of $\approx 10^{-6}$ events per year per galaxy
\cite{NPS,Phi}, in amazing agreement with the GRB event rate
\cite{Piran92}.   It has been suggested \cite{TY} that
many neutron star binaries are born with very close orbits and hence
with very short lifetimes. If this idea is correct, then the merger
rate will be much higher. This will destroy, of course,the nice
agreement between the rates of GRBs  and NS$^2$Ms. Consistency can be
restored if we invoke beaming, which might even be advantageous for
some models.  The short lifetime of those systems, which is the
essence of this idea, makes it impossible to confirm or rule out this
speculation.

NS$^2$Ms result, most likely, in rotating black holes \cite{Dav}. The
process releases $\approx 5 \times 10^{53}$ ergs \cite{CE}.  Most of
this energy escapes as neutrinos and gravitational radiation, but a
small fraction of this energy suffices to power a GRB.  The observed
rate of NS$^2$Ms is similar to the observed rate of GRBs.  This is not
a lot - but this is more than can be said, at present, about any other
GRB model.

\section{Implications to Underground Physics}

Even though GRBs can be detected only by satellites traveling 
outside the atmosphere this phenomenon has several important
implications to other branches of physics and in particular to
underground physics. This is not surprising in view of the unique
character of the fireball model that involves relativistic motion of a
significant amount of particles.  I will discuss some of these
implications now.

It is quite likely that in addition to $\gamma$-rays other particles,
denoted x, are emitted in these events. Let $f_{x-\gamma}$ be the
ratio of energy emitted in these other particles relative to
$\gamma$-rays\footnote{I assume in the following that the
$\gamma$-rays from the GRB and the x particles have the same angular
distribution. This is a reasonable assumption if both are produced by
the fireball's shocks.
It might not be the case if the x particles are
produced by the `''inner engine''. A modification that takes care of
this correction is trivial}. These particles will appear as a
burst accompanying the GRB.  The total fluence of a ``typical'' GRB
observed by BATSE, $F_\gamma$ is $10^{-7}$ergs/cm$^2$, and the fluence
of a ``strong'' burst is about hundred times larger.  Therefore we
should expect accompanying bursts with typical fluences of:
\begin{equation}
F_{x~|prompt} = 0.001 {\rm particles / cm}^2~ f_{x-\gamma}
\end{equation}
$$
\times 
\big({F_\gamma \over 10^7 {\rm ergs/cm}^2}\big) 
\big({E_x \over {\rm GeV}}\big)^{-1} ,
$$ where $E_x$ is the energy of our particles.  This burst will be
spread in time and delayed relative to the GRB if the particles do not
move at the speed of light. Relativistic time delay will be
significant (larger than 10 seconds) if the particles are not massless
and their Lorentz factor is smaller than $10^{8}$!  similarly a
deflection angle of $10^{-8}$ will cause a significant time delay.

In addition to the prompt burst we should expect a continuous
background of these particles.  With one $10^{51}$ergs GRB per $10^6$
years per galaxy we expect $~\sim 10^4$ events per galaxy in a Hubble
time (provided of course that the event rate is constant in
time). This will correspond to a background flux of
\begin{equation}
F_{x~|background} = 3 \cdot 10^{-8} 
{\rm particles /cm^2 sec}~ f_{x-\gamma} 
\end{equation}
$$\times \big ({E_\gamma \over 10^{51} {\rm ergs}}\big) 
\big({R \over 10^{-6} {\rm years/galaxy}}\big)
\big({E_x \over {\rm GeV}}\big)^{-1} .  
$$ For any specific particle that could be produced one should
calculate the ratio $f_{x-\gamma}$ and then compare the expected
fluxes with fluxes from other sources and with the capabilities of
current detectors.

One should distinguish between two types of predictions: (i)
Predictions of the generic fireball model which include low energy
cosmic rays \cite{SP}, UCHERs \cite{Waxman95a,Vietri95} and high
energy neutrinos \cite{Waxman_Bahcall}. (ii) Predictions of specific
models and in particular the NS$^2$M model, which include bursts of
low energy neutrinos \cite{CE} and gravitational waves.

\subsection{Cosmic Rays}

Already in 1990, Shemi \& Piran \cite{SP} pointed out that fireball
model is closely related to Cosmic Rays.  A ``standard'' fireball
model involved the acceleration of $\sim 10^{-7} M_\odot$ of baryons
to a typical energy of 100GeV per baryon.  Protons that leak out of
the fireball will become low energy cosmic rays. However, a comparison
of the GRB rate (one per $10^6$ years per galaxy) with the observed
flux of low energy cosmic rays, suggests that even if $f_{CR-\gamma}
\approx 1$ this will amount only to 1\% to  10\% of the observed cosmic ray
flux at these energies. Cosmic rays are believed to be produced by
SNRs. Since supernovae are ten thousand times more frequent than GRBs,
unless GRBs are much more efficient in producing Cosmic Rays in some
specific energy range their contribution will be swamped by the SNR
contribution.

\subsection{UCHERs - Ultra High Energy Cosmic Rays}

Waxman \cite{Waxman95a} and Vietri \cite{Vietri95} have shown that the
observed flux of UCHERs (above $10^{19}$eV) is consistent with the
idea that these are produced by the fireball shocks provided that
$f_{UCHERs-\gamma} \approx 1$.  SNR cannot not produce such a high
energy particles, while the relativistic shocks of the fireball might
be capable of doing that.  Waxman \cite{Waxman95b} has shown that the
spectrum of UCHERs is  consistent with the expected from Fermi
acceleration within those shocks. An advantage of this source over
other sources is that it is intrinsically optically thin  and the
density of photons at the source that could interact with the UCHERs
and slow them down is rather low. 

\subsection{High Energy Neutrinos}
Waxman and Bahcall \cite{Waxman_Bahcall} suggested that collisions
between protons and photons within the relativistic fireball shocks
produce pions.  These pions produce high energy neutrinos with $E_\nu
\sim 10^{14}$eV and $f_{{\rm high~energy}~\nu-\gamma} > 0.1$.  The
flux of these neutrinos is comparable to the flux of atmospheric
neutrinos but those will be correlated with the position of strong
GRBs. This signal might be detected in future km$^2$ size neutrino
detectors.

\subsection{Gravitational Waves}

If GRBs are associated with NS$^2$Ms then they will be associated with
gravitational waves and low energy neutrinos.  The spiraling in phase
of a NS$^2$M produces a clean chirping gravitational radiation
signal. This signal is the prime target of LIGO and VIRGO, the two
large interferometers that are build now in the USA and in Europe
\cite{Abramovichi}.
The observational scheme of these detectors is heavily dependent on
digging deeply into the noise.  Kochaneck \& Piran
\cite{Kochaneck_Piran} suggested that coincidence with a GRB could
enhance greatly the statistical significance of detection of a
gravitational radiation signal. It will also verify at the same time
this model.

\subsection{Low Energy Neutrinos}
Most of the energy released in a NS$^2$M will be released as low
energy ($\sim 5-10$MeV neutrinos \cite{CE}. The total energy is quite
large $\sim$ a few $ \times 10^{53}$ergs, leading to $f_{{\rm
low~energy}~\nu-\gamma} \approx 100$. However, this neutrino signal
will be quite similar to a supernova neutrino signal - which can be
detected at present only if it is galactic.  Supernovae are ten
thousand times more frequent then GRBs and therefore NS$^2$M neutrinos
constitute an insignificant contribution to the background at this
energy range.

\section{Concluding Remarks}
After thirty years we are finally beginning to understand the nature
of GRBs.  The discovery of the afterglow has demonstrated that we are
on the right track, at least as far as the $\gamma$-ray producing
regions are concerned. This by itself have some fascinating
implications on accompanying UCHER and high energy neutrino signals.
However, we are still uncertain what are the engines that power the
whole phenomenon. My personal impression is that binary neutron
mergers are the best candidates. This model has one specific
prediction - a correlation between GRBs and gravitational radiation
signals. This would confirm or rule out this model next decade when
the next generation of gravitational radiation detectors will begin to
operate.

I thank E. Cohen, J. Katz, S. Kobayashi, R. Narayan, and R. Sari for
helpful discussions. This work was supported by the US-Israel BSF
grant 95-328 and by NASA grant NAG5-3516

\end{document}